\documentstyle[12pt,thmsa,sw20lart]{article}
\input tcilatex
\begin{document}

\title{{\bf Non-local Symmetries of Nonlinear Field Equations: an Algebraic Approach%
}}
\author{M. Leo, R.A. Leo, G. Soliani and P. Tempesta\thanks{%
Address for correspondence: Dott. P. Tempesta, Dipartimento di Fisica,
Universita' degli Studi di Lecce, Via per Arnesano, 73100-Lecce (Italy).
E-mail: Tempesta@le.infn.it.} \\
%EndAName
Dipartimento di Fisica dell'Universit\`{a} di Lecce, 73100 Lecce, Italy.}
\date{}
\maketitle

\begin{abstract}
An algebraic method is devised to look for non-local symmetries of the
pseudopotential type of nonlinear field equations. The method is based on
the use of an infinite-dimensional subalgebra of the prolongation algebra $L$
associated with the equations under consideration. Our approach, which is
applied by way of example to the Dym and the Korteweg-de Vries equations,
allows us to obtain a general formula for the infinitesimal operator of the
non-local symmetries expressed in terms of elements of $L$. The method could
be exploited to investigate the symmetry properties of other nonlinear field
equations possessing nontrivial prolongations.
\end{abstract}

\section{ Introduction}

Local symmetries of differential equations (DEs) are defined by
infinitesimal operators which generally are functions of the independent and
the dependent variables (fields) involved in the equations under
consideration. On the contrary, non-local symmetries are characterized by
infinitesimal operators depending on the global behavior of the fields,
expressed for instance by their integrals [1,2].

The study of {\it non-local} symmetries can be performed following a
procedure which relates these symmetries to the {\it local }symmetries of
certain auxiliary systems of equations connected with the original DEs. An
interesting situation is that where these equations can be expressed as
conservation laws. Indeed, in this case it is possible to introduce new
dependent variables, called potentials, which can be defined by quadratures.
Hence, with a given original system of DEs $\Delta (x,u),\;$where $%
x=(x_{1},...,x_{m})\in R^{m}$ and $u=(u_{1},...,u_{n})\in R^{n},$ one can
associate a system $\Theta (x,u,y)$ (see later) which depends on the set of
potential variables $y$ $=\{y_{i}\}$ as well, with the property that any
solution $(u(x),y(x))$ of $\Theta $ is also a solution of $\Delta .$
Viceversa, for any solution $u(x)$ of $\Delta $ there is a potential $y(x)$
such that the pair $(u(x),y(x))$ is a solution of $\Theta \;$[2]. Of course,
any symmetry group $S_{\Theta }$ of $\Theta $ will be also a symmetry group
of $\Delta .$

More generally, any local group of transformations of $\Theta $ defines a
non-local group of transformations of $\Delta ,$ provided that the
generators of the infinitesimal symmetry transformations (of the independent
and dependent variables) are explicit functions of the potential. Then, the
non-local case can be treated using the same algorhitmic procedures valid
for the local symmetries.

The {\it potential symmetries} can also be exploited to carry out the
symmetry reduction of a given system of differential equations. The case of
partial differential equations has been dealt with by Bluman and Reid [3].
Other kinds of non-local symmetries have been studied in [4-6]. Recently,
Guthrie and Hickman [7] have derived new algebraic structures for the
bi-Hamiltonian version of the Korteweg-de Vries (KdV) equation. Since the
inverse recurrence operator for this equation, because of its bi-Hamiltonian
origin, is still a recurrence operator, it can generate three new families
of generalized symmetries, depending on non-local variables. These
symmetries can be interpreted as isovectors of a prolonged system of the KdV
equation, which are found starting from an infinite-dimensional realization
of the Estabrook-Wahlquist (EW) prolongation algebra.

A remarkable generalization of the concept of non-local symmetry, i.e. the
Lie-B\"{a}cklund (non-local) symmetry, has been devised by Edelen [8] and
Krasil'shchik and Vinogradov [9], where the non-locality character is
carried out by variables of the pseudopotential type. In this context, Galas
has rederived the one-soliton solutions for the KdV and the Dym equations,
and the AKNS system [10]. The non-local Lie-B\"{a}cklund technique presents
the benefit to join up the EW prolongation method [11], which yields the set
of pseudopotentials, to the symmetry reduction approach [1], giving the
tools for analyzing field equations via the group theory. Other interesting
results on the theory of non-local symmetries have been achieved in [12] and
[13].

In this paper we outline a procedure to obtain non-local symmetries of the
psedopotential type of a partial differential equation admitting an
incomplete prolongation algebra $L$ (in the sense that not all of the
commutators of $L$ are known). Equations enjoying this property are
important both from a mathematical point of view and in physical
applications. Our method consists essentially in looking for an
infinite-dimensional subalgebra ${\cal L}$ of $L$ containing as a special
case a finite-dimensional subalgebra ${\cal L}_{o}$ of $L.$ This procedure
is suggested by the fact that in the framework of ${\cal L}_{o}$ only
trivial non-local symmetries can be found. The infinite dimensionality of
the subalgebra seems to be crucial for the determination of nontrivial
non-local symmetries. Within our approach, which is applied to the Dym and
the KdV equations, we obtain a general formula for the infinitesimal
operator of the non-local symmetries expressed in terms of elements of $%
{\cal L}$. In theory, other equations (or systems of equations) endowed with
nontrivial prolongations could be treated in a similar way.

The paper is organized as follows. Section 2 is devoted to some
preliminaries on potential and pseudopotential symmetries. In Sections 3 and
4 we consider the Dym and the KdV equations, respectively. The prolongation
algebras allowed by these equations are incomplete. We apply the algebraic
technique described above to yield the generators of the non-local
symmetries (of the pseudopotential type) together with examples of
interesting solutions of the equations under study. Some of these solutions
are well-known and do not represent, of course, the main goal of the paper,
which is based conversely on a unifying (algebraic) tool to find non-local
symmetries. Finally, Section 5 contains some concluding remarks, while in
the Appendices A and B details of the calculations are reported.

\section{Potential and pseudopotential symmetries}

Let us deal with a system of nonlinear field equations

\smallskip

\begin{equation}
\Delta (x,u)\equiv u_{t}+K(x,t,u,u_{x},...,u_{x...x})=0,  \tag{2.1}
\end{equation}

with $(x,t)\in R^{2},$ $u=(u_{1},...,u_{n})\in R^{n}$ and $u_{t}=D_{t}(u),$ $%
u_{x}=D(u),$ and so on, where $D_{t\text{ }}$and $D_{x}$ stand for the total
derivatives with respect to $t$ and $x.$

Let us suppose that the system (2.1) admits conservation laws of the type

\begin{equation}
\frac{\partial }{\partial t}F_{i}(x,t,u,u_{x},...,u_{x...x})=\frac{\partial 
}{\partial x}G_{i}(x,t,u,u_{x},...,u_{x...x}).  \tag{2.2}
\end{equation}

Equations (2.2) allow us to introduce a set of potentials $y$ $=\{y_{i}\}$
such that

\begin{equation}
y_{i,x}^{{}}=F_{i}^{{}},  \tag{2.3a}
\end{equation}
\begin{equation}
y_{i,t}^{{}}=G_{i}^{{}}.  \tag{2.3b}
\end{equation}
Then one can consider non-local Lie-B\"{a}cklund operators for the system
(2.1) of the form

\begin{equation}
V=Q_{j}^{{}}(x,t,u,u_{x},...,u_{x...x},y)\frac{\partial }{\partial u_{j}^{{}}%
}.  \tag{2.4}
\end{equation}

The non-local character of $V$ is due to the fact that it depends on the
variables $y_{i},$ which are defined by quadratures.

The Lie-B\"{a}cklund non-local symmetries corresponding to $V$ are
equivalent to the Lie-point symmetries of the system $\Theta (x,u,y)$
constituted by Eq. (2.1) and by the set of conservation laws (2.2). These
symmetries are generated by vector fields of the form

\begin{equation}
V_{\Theta }=\xi (x,t,u,y)\frac{\partial }{\partial x}+\tau (x,t,u,y)\frac{%
\partial }{\partial t}+\phi _{j}(x,t,u,y)\frac{\partial }{\partial u_{j}}%
+\pi _{i}(x,t,u,y)\frac{\partial }{\partial y_{i}}  \tag{2.5}
\end{equation}
where the coefficients depend {\it explicitly} on the potential variables.

\smallskip Non-local symmetries can be extended to the case in which the
(non-local) symmetry operators depend on psudopotential variables [10]. In
opposition to the potential variables, pseudopotential variables can be
defined by the set of{\it \ implicit} equations

\begin{equation}
y_{i,x}^{{}}=F_{i}(x,t,u,u_{x},...,u_{x...x},y),\text{ }%
y_{i,t}^{{}}=G_{i}(x,t,u,u_{x},...,u_{x...x},y),  \tag{2.6}
\end{equation}

\smallskip

where the functional dependence of $F_{i}$ and $G_{i}$ includes the
pseudopotential variables also. The compatibility condition $%
y_{i,xt}=y_{i,tx}$ reproduces the original equations of the type (2.1).\
Equations (2.6) allow us to obtain, in theory, the prolongation algebra
associated with the differential equations under consideration and the
related spectral problem [12].

In the following we shall display two case studies concerning the
determination of non-local symmetries based on the use of pseudopotentials.

\smallskip

\section{Non-local symmetries of the Dym equation}

The EW prolongation technique applied to the Dym equation

\smallskip

\begin{equation}
u_{t}=u^{3}u_{xxx}  \tag{3.1}
\end{equation}

gives

\begin{equation}
y_{i,x}=F_{i}=\frac{A_{i}}{u^{2}}+B_{i},  \tag{3.2a}
\end{equation}

\begin{equation}
y_{i,t}=G_{i}=-2A_{i}u_{xx}-2C_{i}u_{x}-\frac{2}{u}[A,C]_{i}+2u[B,C]_{i}, 
\tag{3.2b}
\end{equation}

where $A_{i},B_{i},C_{i}$ are functions depending on the pseudopotential
variables only, and $[A,C]_{i},[B,C]_{i}$ are Lie brackets defined by

\[
\lbrack A,C]_{i}=A_{j}\frac{\partial }{\partial y_{j}}C_{i}-C_{j}\frac{%
\partial }{\partial y_{j}}A_{i}, 
\]

and so on.

Hereafter, for simplicity we shall omit the index $i$ and adopt an operator
formalism, in the sense that we shall define the operators

\[
\stackrel{}{\stackrel{\wedge }{F}=F_{j}}\frac{\partial }{\partial y_{j}^{{}}}%
,\stackrel{\wedge }{G}=\stackrel{}{G_{j}}\frac{\partial }{\partial y_{j}^{{}}%
},\stackrel{\wedge }{A}=A_{j}\frac{\partial }{\partial y_{j}^{{}}},..., 
\]

etc.. Then the Lie brackets are transformed into commutators, viz.

$[\stackrel{\wedge }{A},\stackrel{\wedge }{C}]\equiv [A,C]_{j}^{{}}\frac{%
\partial }{\partial y_{j}^{{}}},$ ... . Furthermore, to avoid redundance of
symbols, we shall continue to use $F$ instead of $\stackrel{\wedge }{F},$
and so on.

\smallskip From the compatibility condition $y_{xt}=y_{tx}$ we have that the
operators $A,B,C$ obey the commutation rules

\begin{equation}
\lbrack A,B]=C,  \tag{3.3a}
\end{equation}

\begin{equation}
\lbrack A,[A,C]]=0,  \tag{3.3b}
\end{equation}

\begin{equation}
\lbrack B,[B,C]]=0.  \tag{3.3c}
\end{equation}

\smallskip Equations (3.3a)-(3.3c) define an incomplete Lie algebra. In
order to investigate the existence of non-local symmetries (of the
pseudopotential type) of Eq. (3.1), let us deal with the infinitesimal
transformation 
\begin{equation}
\stackrel{}{\stackrel{\sim }{u}=u+}\varepsilon \varphi (u,u_{x},y), 
\tag{3.4}
\end{equation}

where $\varphi $ is a function to be found and $\varepsilon $ is a real
parameter.

By imposing (3.4), Eq. (3.1) yields

\begin{equation}
\varphi _{t}=3u^{2}\varphi u_{xxx}+u^{3}\varphi _{xxx},  \tag{3.5}
\end{equation}

\smallskip where $\varphi _{t}=D_{t}\varphi $ and $\varphi _{x}=D_{x}\varphi
.$

At this stage it is convenient to adopt a procedure which consists in the
use of the prolongation algebra (3.3a)-(3.3c) without employing any specific
representation for the vector fields $A,B,C.$

By expliciting (3.5) and equating to zero the coefficients of the
derivatives of $u$ regarded as independent functions, we get

\begin{equation}
\varphi =(u_{x}-uB)\varphi _{o}(y),  \tag{3.6}
\end{equation}

where $\varphi _{o}$ is a function of $y$ satisfying the constraints

\begin{equation}
A\varphi _{o}=0,  \tag{3.7a}
\end{equation}
\begin{equation}
A^{2}B\varphi _{o}=0,  \tag{3.7b}
\end{equation}
\begin{equation}
B^{3}\varphi _{o}=0,  \tag{3.7c}
\end{equation}
\begin{equation}
BCB\varphi _{o}=0,  \tag{3.7d}
\end{equation}
\begin{equation}
2[A,C]B\varphi _{o}+A^{2}B^{2}\varphi _{o}+ABAB\varphi _{o}=0.  \tag{3.7e}
\end{equation}
Equations (3.7a) and (3.7b) entail 
\begin{equation}
\lbrack A,C]\varphi _{o}=0.  \tag{3.8}
\end{equation}

Now let us introduce an infinitesimal transformation for the pseudopotential 
$y$ analogous to (3.4):

\begin{equation}
\stackrel{\sim }{y}=y+\varepsilon \eta (u,y),  \tag{3.9}
\end{equation}

where $\eta (u,y)$ is a vector field to be determined.

We observe that both $\eta $ and $\varphi $ could have a functional
dependence more complicated than that made here; indeed, the choice
expressed by Eqs. (3.4) and (3.9) corresponds to a minimal assumption. We
notice also that if we take $\varphi $ independent from $u_{x},$ only the
identity arises as a symmetry.

Combining together (3.9) and (3.2a) we obtain

\[
\stackrel{\sim }{y}_{x}=y_{x}+\varepsilon \eta _{x}(u,y)=F(y+\varepsilon
\eta ,u+\varepsilon \varphi )=F+\varepsilon (F_{y}\eta +F_{u}\varphi ). 
\]
Hence

\[
u_{x}\eta _{u}+[F,\eta ]-\varphi F_{u}=0, 
\]

namely

\begin{equation}
u_{x}\eta _{u}+\frac{1}{u^{2}}[A,\eta ]+[B,\eta ]+\frac{2}{u^{3}}%
u_{x}\varphi \Sb o  \\  \endSb A-\frac{2}{u^{2}}(B\varphi _{o})A=0. 
\tag{3.10}
\end{equation}
Taking equal to zero the coefficients of the independent functions of $u$
and their partial derivatives appearing in (3.10), we find

\begin{equation}
\eta =\frac{1}{u^{2}}\varphi _{o}A+\eta _{o},  \tag{3.11}
\end{equation}
$\eta _{o}=\eta _{o}(y)$ being a vector field of integration, and

\begin{equation}
\lbrack A,\varphi _{o}A]=0,  \tag{3.12a}
\end{equation}
\begin{equation}
\lbrack A,\eta _{o}]+[B,\varphi _{o}A]-2(B\varphi _{o})A=0,  \tag{3.12b}
\end{equation}
\begin{equation}
\lbrack B,\eta _{0}]=0.  \tag{3.12c}
\end{equation}
In a similar manner, starting from (3.2b) and taking account of (3.9), we
obtain (at the first order in $\varepsilon )$:

\smallskip 
\begin{equation}
u^{3}u_{xxx}\eta _{u}+[G,\eta ]-\frac{2}{u^{2}}\varphi [A,C]-2\varphi
[B,C]+2\varphi _{x}C+2\varphi _{xx}A=0.  \tag{3.13}
\end{equation}
Equation (3.13) can be explicited to give

\begin{equation}
(AB\varphi _{o})A-[C,\varphi _{o}A]-\varphi _{o}[A,C]=0,  \tag{3.14a}
\end{equation}
\begin{equation}
\lbrack C,\eta _{o}]+\varphi _{o}[B,C]+(B^{2}\varphi _{o})A=0,  \tag{3.14b}
\end{equation}
\begin{equation}
\lbrack [A,C],\varphi _{o}A]=0,  \tag{3.14c}
\end{equation}
\begin{equation}
\lbrack [B,C],\eta _{o}]+(B\varphi _{o})[B,C]-(B^{2}\varphi _{o})C=0, 
\tag{3.14d}
\end{equation}
\begin{equation}
\begin{array}{l}
\lbrack [B,C],\varphi _{o}A]-[[A,C],\eta _{o}]+(B\varphi
_{o})[A,C]-A(B\varphi _{o})C-(AB^{2}\varphi _{o})A \\ 
\qquad \qquad \qquad \qquad \qquad -(BAB\varphi _{o})A=0.
\end{array}
\tag{3.14e}
\end{equation}

Finally, the infinitesimal generator of the non--local symmetries of the
pseudopotential type for the Dym equation is (see (3.6) and (3.11)) 
\begin{equation}
V_{NL}=\varphi \partial _{u}+\eta =(u_{x}\varphi _{o}-uB\varphi
_{o})\partial _{u}+\frac{1}{u^{2}}\varphi _{o}A+\eta _{o}.  \tag{3.15}
\end{equation}

The problem of the determination of the non-local symmetries of the Dym
equation is led to find suitable representations of the incomplete
prolongation algebra (3.3a)-(3.3c) and the constraints associated with this
equation.

\smallskip In this context, we shall show below that by choosing a
finite-dimensional representation of the prolongation algebra (3.3a)-(3.3c)
of the $sl(2,R)$ type, we arrive at trivial transformations only.

To this aim, let us set

\begin{equation}
\lbrack A,C]=b_{1}A+b_{2}B+b_{3}C,  \tag{3.16a}
\end{equation}
\begin{equation}
\lbrack B,C]=c_{1}A+c_{2}B+c_{3}C,  \tag{3.16b}
\end{equation}
\begin{equation}
\lbrack A,B]=C,  \tag{3.16c}
\end{equation}

\smallskip where $b_{j},c_{j}$ are constants.

Then, from (3.3a) and (3.3c) we find $b_{2}=b_{3}=0$ and $c_{1}=c_{3}=0,$
respectively. Thus, putting $b_{1}=-2\lambda ,$ $b_{2}=2\lambda $ Eqs.
(3.16a)-(3.16c) become

\begin{equation}
\lbrack A,C]=-2\lambda A,  \tag{3.17a}
\end{equation}

\begin{equation}
\lbrack B,C]=2\lambda B,  \tag{3.17b}
\end{equation}
\begin{equation}
\lbrack A,B]=C.  \tag{3.17c}
\end{equation}
On the other hand, Eq. (3.14d) gives

\begin{equation}
2\lambda (B\varphi _{o})B=(B^{2}\varphi _{o})C,  \tag{3.18}
\end{equation}
while from (3.14e):

\begin{equation}
(2\lambda (B\varphi _{o})-(AB^{2}\varphi _{o})-(BA(B\varphi
_{o}))A=(AB\varphi _{o})C.  \tag{3.19}
\end{equation}
Applying (3.19) to $\varphi _{o}$ yields

\[
(AB\varphi _{o})(C\phi _{o})=0, 
\]

\smallskip from which

\smallskip

\begin{equation}
C\varphi _{o}=0.  \tag{3.20}
\end{equation}

As a consequence, the constraint (3.18) provides $B\varphi _{o}=0.$ To
conclude, the quotient algebra (3.17a)-(3.17c) leads to the relations

\begin{equation}
A\varphi _{o}=0,B\varphi _{o}=0,C\varphi _{o}=0,  \tag{3.21}
\end{equation}

which tell us that $\varphi _{o}=const.$

\subsection{\protect\smallskip {\it The method of the ''extended'' algebra}}

\bigskip We have seen that within a finite-dimensional subalgebra (such as $%
sl(2,R)$) of the prolongation algebra $L$ defined by (3.3a)-(3.3c), we have
not been able to find nontrivial non-local symmetries. Then, we have
exploited an infinite-dimensional subalgebra of $L$. As we show below, this
approach succeeds. To this aim, let us introduce the operators

\begin{equation}
A=A_{0}+\varphi _{0}A_{1},  \tag{3.22}
\end{equation}

\begin{equation}
B=B_{o},  \tag{3.23}
\end{equation}

\begin{equation}
C=C_{0}-(B_{0}\varphi _{0})A_{1},  \tag{3.24}
\end{equation}

into the commutation relations (3.3a)-(3.3c), where

\begin{equation}
\lbrack A_{0},B_{0}]=C_{0},[A_{0,}C_{0}]=-2\lambda
A_{0},[B_{0},C_{0}]=2\lambda B_{0}  \tag{3.25}
\end{equation}

and $A_{1}$ denotes an operator obeying the commutation rules

\begin{equation}
\lbrack A_{1},A_{0}]=0,[A_{1},B_{0}]=0,[A_{1},C_{0}]=0.  \tag{3.26}
\end{equation}

Furthermore, we assume that

\begin{equation}
A_{1}\varphi _{0}=0,  \tag{3.27}
\end{equation}

and

\begin{equation}
A_{0}\varphi _{0}=A_{o}^{2}B_{o}\varphi _{o}=B_{o}^{3}\varphi _{o}=0. 
\tag{3.28}
\end{equation}

(We observe that the constraints (3.28) are really special cases of (3.7a),
(3.7b) and (3.7c), respectively).

By virtue of (3.22)-(3.28), we can prove directly that the commutators

$\smallskip $%
\begin{equation}
\lbrack A,B_{0}]=C_{0}-(B_{0}\varphi _{0})A_{1},  \tag{3.29a}
\end{equation}
\ 

\begin{equation}
\lbrack A,C]=-2\lambda A_{0}-\{(A_{0}B_{0}\varphi _{0})+(C_{0}\varphi
_{0})\}A_{1},  \tag{3.29b}
\end{equation}

\begin{equation}
\lbrack B_{0},C]=2\lambda B_{0}-(B_{0}^{2}\varphi _{0})A_{1},  \tag{3.29c}
\end{equation}

realize the prolongation algebra ${\cal L}$ expressed by (3.3a)-(3.3c).

When \ $A_{1}=0,\;$we obtain $A=A_{0},C=C_{0},\;$and\ the commutation rules
(3.29a)-(3.29c) reproduce just the $sl(2,R)\;$algebra (3.25). In some sense,
the algebra defined by Eqs. (3.29a)-(3.29c) plays the role of an
''extended'' algebra, ${\cal L}_{E}$ , relatively to the $sl(2,R)$ algebra
(3.25). Since $A_{1}$ is multiplied by an arbitrary function, it turns out
that ${\cal L}_{E}\;$is an infinite-dimensional$\;$subalgebra of ${\cal L}$
(satisfying all the constraints involved by the theory of non-local
symmetries). In the following, we shall see that the use of ${\cal L}_{E}$
instead of $sl(2,R)$ enables us to obtain nontrivial non-local symmetries.

Now let us demand that

\begin{equation}
\lbrack A_{o},\eta _{o}]=0,[B_{o},\eta _{o}]=0,[C_{o},\eta _{o}]=0. 
\tag{3.30}
\end{equation}

A possible bidimensional realization of the algebra (3.25) is

\smallskip

\begin{equation}
A_{o}=\lambda \partial _{y_{1}},  \tag{3.31a}
\end{equation}
\begin{equation}
B_{o}=-y_{1}^{2}\partial _{y_{1}}+y_{1}\partial _{y_{2}},  \tag{3.31b}
\end{equation}
\begin{equation}
C_{o}=-2\lambda y_{1}\partial _{y_{1}}+\lambda \partial _{y_{2}}. 
\tag{3.31c}
\end{equation}
From (3.12b) we find

\begin{equation}
\lbrack A_{1},\eta _{o}]=\frac{\eta _{o}\varphi _{o}}{\varphi _{o}}A_{1}+%
\frac{B_{o}\varphi _{o}}{\varphi _{o}}A_{o}+C_{o},  \tag{3.32}
\end{equation}
while (3.14b) gives

\begin{equation}
\lbrack A_{1},\eta _{o}]=\frac{(\eta _{o}B_{o}\varphi _{o})}{(B_{o}\varphi
_{o})}A_{1}^{{}}+\frac{2\lambda \varphi _{o}B_{o}}{(B_{o}\varphi _{o})}+%
\frac{(B_{o}^{2}\varphi _{o})A_{o}}{(B_{o}\varphi _{o})}.  \tag{3.33}
\end{equation}
Furthermore, from (3.14d) we deduce

\begin{equation}
\lbrack A_{1},\eta _{o}]=\frac{(\eta _{o}B_{o}^{2}\varphi _{o})}{%
(B_{o}^{2}\varphi _{o})}A_{1}^{{}}+\frac{2\lambda \varphi _{o}B_{o}}{%
(B_{o}^{2}\varphi _{o})}-C_{o}.  \tag{3.34}
\end{equation}
By comparing (3.32), (3.33) and (3.34) we have

\begin{equation}
\frac{(\eta _{o}\varphi _{o})}{\varphi _{o}}=\frac{(\eta _{o}B_{o}\varphi
_{o})}{(B_{o}\varphi _{o})}=\frac{(\eta _{o}B_{o}^{2}\varphi _{o})}{%
(B_{o}^{2}\varphi _{o})},  \tag{3.35}
\end{equation}
\begin{equation}
\frac{(B_{o}\varphi _{o})}{\varphi _{o}}A_{o}+C_{o}=\frac{2\lambda \varphi
_{o}B_{o}}{(B_{o}\varphi _{o})}+\frac{(B_{o}^{2}\varphi _{o})A_{o}}{%
(B_{o}\varphi _{o})}=\frac{2\lambda (B_{o}\varphi _{o})}{(B_{o}^{2}\varphi
_{o})}B_{o}-C_{o}.  \tag{3.36}
\end{equation}
Now we are ready to derive $\varphi _{o},A_{1},$ and $\eta _{o}.$ In doing
so, let us suppose that these quantities can be expressed as

\begin{equation}
\varphi _{o}=\varphi _{o}(y)\equiv \varphi _{o}(y_{1},y_{2},y_{3}), 
\tag{3.37}
\end{equation}
\begin{equation}
A_{1}=a_{1}(y)\partial _{y_{1}}+a_{2}(y)\partial _{y_{2}}+a_{3}(y)\partial
_{y_{3}},  \tag{3.38}
\end{equation}
\begin{equation}
\eta _{o}=f_{1}(y)\partial _{y_{1}}+f_{2}(y)\partial
_{y_{2}}+f_{3}(y)\partial _{y_{3}},  \tag{3.39}
\end{equation}
where $a_{j}$, $f_{j}$ are functions of\ $y\equiv (y_{1},y_{2},y_{3})$ to be
determined.

From (3.36) we deduce

\begin{equation}
B_{o}\varphi _{o}=2\varphi _{o}y_{1},  \tag{3.40}
\end{equation}
which yields

\begin{equation}
\varphi _{o}=\phi (y_{3})e^{2y_{2}},  \tag{3.41}
\end{equation}
$\phi (y_{3})$ being a function of integration depending on $y_{3}$ only.

Then, taking account of (3.40), it turns out that (3.35) and (3.36) are
identically satisfied.

On the other hand, since $A_{1}\varphi _{o}=0$ we get (see (3.41) and
(3.38)):

\begin{equation}
2a_{2}(y)\phi (y_{3})+a_{3}(y)\phi _{y_{3}}(y_{3})=0.  \tag{3.42}
\end{equation}
Then, resorting to the commutation rule $[A_{1},B_{o}]=0,$ we can write

\begin{equation}
\lbrack B_{o},\text{ }A_{1}]=(-y_{1}^{2}\partial _{y_{1}}+y_{1}\partial
_{y_{2}})A_{1}\varphi _{o}+A_{1}(y_{1}^{2}\partial _{y_{1}}-y_{1}\partial
_{y_{2}})\varphi _{o}=0,  \tag{3.43}
\end{equation}
which gives $a_{1}(y)=0$.

Furthermore, from $[A_{1},A_{o}]=0$ we infer that the functions $a_{2}$ and $%
a_{3}$ appearing in (3.42) are independent from the pseudopotential variable 
$y_{1},$ i.e. $a_{2}=a_{2}(y_{2},y_{3}),a_{3}=a_{3}(y_{2},y_{3}).$

Finally, the commutation relation $[A_{1},C_{o}]=0$ yields

\begin{equation}
A_{1}=a_{2}(y_{3})\partial _{y_{2}}+a_{3}(y_{3})\partial _{y_{3}}, 
\tag{3.44}
\end{equation}

where $a_{2}$ and $a_{3}$ are functions of the pseudopotential variable $%
y_{3}$ only.

At this point let us consider the operator $\eta _{o},$ which obeys the
commutation rules $[A_{o},\eta _{o}]=0,\;[B_{o},\eta _{o}]=0,\;[C_{o},\eta
_{o}]=0.$ The role of\ $\eta _{o}$ is formally analogous to that played by $%
A_{1}.$ Therefore, we easily get

\begin{equation}
\eta _{o}=f_{2}(y_{3})\partial _{y_{2}}+f_{3}(y_{3})\partial _{y_{3}}. 
\tag{3.45}
\end{equation}

Then, we can exploit the commutator $[A_{1},\eta _{o}]$ expressed by (3.32).
Indeed, substituting (3.44) and (3.45) into (3.32) and equating the
coefficients of $\partial _{y_{2}}$ and $\partial _{y_{3}}$ to zero, we are
led to the relations

\begin{equation}
f_{3}a_{2y_{3}}-a_{3}f_{2y_{3}}+2f_{2}a_{2}+f_{3}a_{2}\frac{\phi _{y_{3}}}{%
\phi }+\lambda =0,  \tag{3.46}
\end{equation}

\begin{equation}
-a_{3}f_{3y_{3}}+a_{3y_{3}}f_{3}+2f_{2}a_{3}+f_{3}a_{3}\frac{\phi _{y_{3}}}{%
\phi }=0.  \tag{3.47}
\end{equation}
We remark that the same relations come from (3.14e).

Equations (3.42), (3.46) and (3.47) represent an overdetermined system with
unknowns $a_{2},a_{3},f_{2},f_{3},$ and $\phi .$ The knowledge of these
quantities provides the function $\varphi _{o}$ and the operators $%
A_{1},\eta _{o}$.

\smallskip By using the realization (3.31a)-(3.31c), (3.44) and (3.45), the
generator (3.15) takes the form

\begin{equation}
\begin{array}{l}
V_{NL}=-\phi (y_{3})e^{2y_{2}}\partial _{x}-2u\phi
(y_{3})y_{1}e^{2y_{2}}\partial _{u}+\phi (y_{3})e^{2y_{2}}y_{1}^{2}\partial
_{y_{1}} \\ 
\qquad \qquad +[f_{2}(y_{3})-y_{1}\phi (y_{3})e^{2y_{2}}]\partial
_{y_{2}}+f_{3}(y_{3})\partial _{y_{3}}.
\end{array}
\tag{3.48}
\end{equation}

Thus, the corresponding group transformations arise from the differential
equations

\smallskip 
\begin{equation}
\frac{d\stackrel{\sim }{x}}{d\varepsilon }=-\phi (\stackrel{\sim }{y}%
_{3})e^{2\stackrel{\sim }{y}_{2}},  \tag{3.49a}
\end{equation}

\begin{equation}
\frac{d\stackrel{\sim }{u}}{d\varepsilon }=-2\stackrel{\sim }{u}\phi (%
\stackrel{\sim }{y}_{3})\stackrel{\sim }{y_{1}}e^{2\stackrel{\sim }{y}_{2}},
\tag{3.49b}
\end{equation}
\begin{equation}
\frac{d\stackrel{\sim }{y_{1}}}{d\varepsilon }=\phi (\stackrel{\sim }{y}_{3})%
\stackrel{}{\stackrel{\sim }{y}_{1}^{2}}e^{2\stackrel{\sim }{y}_{2}}, 
\tag{3.49c}
\end{equation}
\begin{equation}
\frac{d\stackrel{\sim }{y_{2}}}{d\varepsilon }=-\phi (\stackrel{\sim }{y}%
_{3})\stackrel{\sim }{y}_{1}e^{2\stackrel{\sim }{y}_{2}}+f_{2}(\stackrel{%
\sim }{y}_{3}),  \tag{3.49d}
\end{equation}
\begin{equation}
\frac{d\stackrel{\sim }{y_{3}}}{d\varepsilon }=f_{3}(\stackrel{\sim }{y}%
_{3}),  \tag{3.49e}
\end{equation}
where $\stackrel{\sim }{t}=t,\varepsilon \;$is the group parameter, and the
boundary conditions

\begin{equation}
\stackrel{\sim }{x}\mid _{\varepsilon =0}=x,\stackrel{\sim }{u}\mid
_{\varepsilon =0}=u,\stackrel{\sim }{y_{1}}\mid _{\varepsilon =0}=y_{1},%
\stackrel{\sim }{y_{2}}\mid _{\varepsilon =0}=y_{2},\stackrel{\sim }{y_{3}}%
\mid _{\varepsilon =0}=y_{3},  \tag{3.50}
\end{equation}
are considered.

Now, in order to illustrate how our method works, let us consider the
trivial solution $u=-1$ to Eq. (3.1). Consequently, the prolongation
equations (3.2a) and (3.2b) provide

\smallskip

\begin{equation}
y_{1x}=\lambda -y_{1}^{2},  \tag{3.51a}
\end{equation}

\begin{equation}
y_{1t}=-4\lambda ^{2}+4\lambda y_{1}^{2},  \tag{3.51b}
\end{equation}

\begin{equation}
y_{2x}=y_{1}+a_{2}\phi (y_{3})e^{2y_{2}},  \tag{3.51c}
\end{equation}
\begin{equation}
y_{2t}=-8\lambda a_{2}\phi (y_{3})e^{2y_{2}}-4\lambda
y_{1}^{{}}+4a_{2}y_{1}^{2}\phi (y_{3})e^{2y_{2}},  \tag{3.51d}
\end{equation}

\begin{equation}
y_{3x}=a_{3}\phi (y_{3})e^{2y_{2}},  \tag{3.51e}
\end{equation}
\begin{equation}
y_{3t}=-8\lambda a_{3}\phi (y_{3})e^{2y_{2}}+4a_{3}y_{1}^{2}\phi
(y_{3})e^{2y_{2}},  \tag{3.51f}
\end{equation}
where here $y_{1},y_{2},y_{3}$ have not to be regarded as vector fields, but
functions of $(x,t$).

After some manipulations, from the equations (3.49a)-(3.49e) and
(3.51a)-(3.51f) we obtain

\begin{equation}
\stackrel{\sim }{u}=u\frac{y_{1}^{2}}{\stackrel{\sim }{y}_{1}^{2}} 
\tag{3.52}
\end{equation}

where $y_{1}$ can be derived by solving the pair of Riccati equations
(3.51a) and (3.51b). By choosing for example $\lambda >0,$ we obtain

\smallskip

\begin{equation}
y_{1}=\sqrt{\lambda }\frac{e^{\sqrt{\lambda }\xi }-ae^{-\sqrt{\lambda }\xi }%
}{e^{\sqrt{\lambda }\xi }+ae^{-\sqrt{\lambda }\xi }},  \tag{3.53}
\end{equation}

\smallskip

where $a$ is a constant of integration, and $\xi =x-4\lambda t.$

By scrutinizing the remaining equations involving the prolongation
variables, i.e. (3.51c)-(3.51f), we have

\begin{equation}
\stackrel{\sim }{y}_{1}=y_{1}\frac{1-2\varepsilon \chi \lambda a(x-12\lambda
t)+\frac{\varepsilon \chi \sqrt{\lambda }}{2}(e^{2\sqrt{\lambda }\xi
}-a^{2}e^{-2\sqrt{\lambda }\xi })}{1-2\varepsilon \chi \lambda a(x-12\lambda
t)-\frac{\varepsilon \chi \sqrt{\lambda }}{2}(e^{2\sqrt{\lambda }\xi
}-a^{2}e^{-2\sqrt{\lambda }\xi })}.  \tag{3.54}
\end{equation}

\smallskip

In order to look for interesting solutions to Eq. (3.1) (starting from the
trivial solution $u=-1),$ we have to use Eq. (3.52) and write the variables $%
x,t$ in terms of $\stackrel{\sim }{x},t.$ In doing so, since $t=\stackrel{%
\sim }{t},$ it is sufficient to consider Eqs. (3.49a) and (3.49c). These
yield

\smallskip

\begin{equation}
\stackrel{\sim }{x}=x+\frac{1}{\stackrel{\sim }{y_{1}}}-\frac{1}{y_{1}}. 
\tag{3.55}
\end{equation}

\smallskip

Then, by using (3.54) and (3.53) we have ($t=\stackrel{\sim }{t})$

\smallskip

\begin{equation}
\stackrel{\sim }{x}=x-\frac{\varepsilon \chi (e^{\sqrt{\lambda }\xi }+ae^{-%
\sqrt{\lambda }\xi })^{2}}{1-2\varepsilon \chi \lambda a(x-12\lambda t)+%
\frac{\varepsilon \chi \sqrt{\lambda }}{2}(e^{2\sqrt{\lambda }\xi
}-a^{2}e^{-2\sqrt{\lambda }\xi })}.  \tag{3.56}
\end{equation}

\smallskip From (3.56) we can derive, formally, $x$ as a function of $\;%
\stackrel{\sim }{x}\;$and $\stackrel{\sim }{t}$: $x=\tau (\stackrel{\sim }{x}%
,\stackrel{\sim }{t}).$ Then, from Eq. (3.52) (with $u=-1)$ we obtain

\smallskip 
\begin{equation}
\stackrel{\sim }{u}=-\left\{ \frac{1-2\varepsilon \chi \lambda a[\stackrel{}{%
\tau (\stackrel{\sim }{x},\stackrel{\sim }{t})}-12\lambda \stackrel{\sim }{t}%
]-\frac{\varepsilon \chi \sqrt{\lambda }}{2}(e^{2\sqrt{\lambda }\stackrel{%
\sim }{\xi }}-a^{2}e^{-2\sqrt{\lambda }\stackrel{\sim }{\xi }})}{%
1-2\varepsilon \chi \lambda a[\stackrel{}{\tau (\stackrel{\sim }{x},%
\stackrel{\sim }{t})}-12\lambda \stackrel{\sim }{t}]+\frac{\varepsilon \chi 
\sqrt{\lambda }}{2}(e^{2\sqrt{\lambda }\stackrel{\sim }{\xi }}-a^{2}e^{-2%
\sqrt{\lambda }\stackrel{\sim }{\xi }})}\right\} ^{2}  \tag{3.57}
\end{equation}

\smallskip

where $\xi =\tau (\stackrel{\sim }{x},\stackrel{\sim }{t})-4\lambda 
\stackrel{\sim }{t}.$

\smallskip We remark that for $a=0,$ $\chi =-1$ and $\varepsilon <0,$ Eq.
(3.57) produces the (formal) solitary wave solution

\begin{equation}
\stackrel{\sim }{u}=\sec h^{2}\sqrt{\lambda }(\stackrel{\sim }{\xi }+\delta
)-1,  \tag{3.58}
\end{equation}

\smallskip

where $\delta $ is a constant defined by $\frac{\mid \varepsilon \mid \sqrt{%
\lambda }}{2}=e^{2\sqrt{\lambda }\delta }.$ This solution corresponds to
that found in [10]. By choosing in the Riccati equations (3.51a) and (3.51b) 
$\lambda <0,$ a procedure similar to that employed to derive formula (3.57)
leads to the solution (A3) reported in Appendix A.

To conclude this Section, we observe that the solutions (3.57) and (A3) are
a consequence of the choice $u=-1$ (i.e. the trivial solution of the Dym
equation) in the B\"{a}cklund transformation (3.52). Of course, in theory
other choices should be possible and, correspondingly, other solutions
should be derived.

\section{Non-local symmetries of the Korteweg-de Vries equation}

Another interesting case which constitutes a good laboratory for checking
the validity of our approach to non-local symmetries with pseudopotentials,
is given by the KdV equation

\begin{equation}
u_{t}+6uu_{x}+u_{xxx}=0.  \tag{4.1}
\end{equation}

The pseudopotential for Eq. (4.1) is defined by

\begin{equation}
y_{x}=F=\frac{1}{2}u^{2}A+uB+C,  \tag{4.2a}
\end{equation}

\begin{equation}
\begin{array}{l}
y_{t}=G=(-uu_{xx}+\frac{1}{2}u_{x}^{2}-2u^{3})A-(u_{xx}+3u^{2})B-\frac{1}{2}%
u^{2}[B,D] \\ 
\qquad \qquad \qquad \qquad -u[C,D]+u_{x}D+E,
\end{array}
\tag{4.2b}
\end{equation}

where $A,B,C,D,E$ are vector fields (depending on $y\equiv \{y_{j}\}$ only)
satisfying the commutation relations

\smallskip

\begin{equation}
\lbrack A,B]=[A,C]=[A,D]=0,  \tag{4.3a}
\end{equation}

\begin{equation}
\lbrack A,[B,D]]=0,  \tag{4.3b}
\end{equation}

\begin{equation}
\lbrack B,[B,D]]=[A,[D,C]]=0,  \tag{4.3c}
\end{equation}

\begin{equation}
\lbrack A,E]=3[C,[B,D]]+6D,  \tag{4.3d}
\end{equation}

\begin{equation}
\lbrack B,E]=[C,[C,D]],  \tag{4.3e}
\end{equation}

\begin{equation}
\lbrack C,E]=0,  \tag{4.3f}
\end{equation}

\begin{equation}
\lbrack C,[B,D]]=[B,[C,D]],  \tag{4.3g}
\end{equation}

\begin{equation}
D=[C,B].  \tag{4.3h}
\end{equation}

The prolongation algebra (4.3a)-(4.3h) is incomplete.

Now let us carry out the infinitesimal transformation

\begin{equation}
\stackrel{\sim }{u}=u+\varepsilon \varphi (u,y).  \tag{4.4}
\end{equation}
Then, Eq. (4.1) provides

\begin{equation}
\varphi _{t}+6u\varphi _{x}+6\varphi u_{x}+\varphi _{xxx}=0.  \tag{4.5}
\end{equation}
Equation (4.5) can be managed to get the following set of constraints:

\smallskip

\begin{equation}
A\varphi =0,  \tag{4.6a}
\end{equation}
\begin{equation}
B^{2}\varphi =0,  \tag{4.6b}
\end{equation}
\begin{equation}
CB\varphi =-2\varphi ,  \tag{4.6c}
\end{equation}
\begin{equation}
-3B\varphi -\frac{1}{2}[B,D]\varphi +6B\varphi +B^{2}C\varphi +BCB\varphi =0,
\tag{4.6d}
\end{equation}

\begin{equation}
-[C,D]\varphi +6C\varphi +BC^{2}\varphi +CBC\varphi +C^{2}B\varphi =0, 
\tag{4.6e}
\end{equation}

\begin{equation}
E\varphi +C^{3}\varphi =0,  \tag{4.6f}
\end{equation}

where $\varphi =\varphi (y),$ i.e. $\varphi $ depends on the pseudopotential
only.

Now let us perform the infinitesimal transformation for the pseudopotential $%
y:$ 
\begin{equation}
\stackrel{\sim }{y}=y+\varepsilon \eta (u,y).  \tag{4.7}
\end{equation}

By virtue of (4.7), Eq. (4.2a) yields

\begin{equation}
u_{x}\eta _{u}+[F,\eta ]-\varphi F_{u}=0.  \tag{4.8}
\end{equation}

\smallskip Equation (4.8) tells us that $\eta _{u}=0,$ i.e. $\eta =\eta (y).$
Moreover, the following conditions

\smallskip

\begin{equation}
\lbrack B,\eta ]=0,  \tag{4.9a}
\end{equation}

\begin{equation}
\lbrack C,\eta ]-\varphi B=0,  \tag{4.9b}
\end{equation}

\begin{equation}
\lbrack D,\eta ]+(B\varphi )B=0,  \tag{4.9c}
\end{equation}

\begin{equation}
\lbrack [B,D],\eta ]=0,  \tag{4.9d}
\end{equation}
\begin{equation}
-[[C,D],\eta ]+6\varphi B+\varphi [B,D]-(B\varphi )D+(BC\varphi
)B+(CB\varphi )B=0,  \tag{4.9e}
\end{equation}

\begin{equation}
\lbrack E,\eta ]+\varphi [C,D]-(C\varphi )D+(C^{2}\varphi )B=0,  \tag{4.9f}
\end{equation}

\smallskip

hold.

Then, our task is to find the infinitesimal generator of the non-local
symmetries for Eq. (4.1), which reads

\smallskip 
\begin{equation}
V_{NL}=\varphi \partial _{u}+\eta .  \tag{4.10}
\end{equation}

\smallskip

To this aim, first we look for a finite-dimensional representation of the
prolongation algebra (4.3a)-(4.3h). Assuming $A=0,$ a representation of this
kind is

\smallskip 
\begin{equation}
\lbrack C_{o},B_{o}]=D_{o},  \tag{4.11a}
\end{equation}
\begin{equation}
\lbrack B_{o},D_{o}]=-2B_{o},  \tag{4.11b}
\end{equation}
\begin{equation}
\lbrack C_{o},D_{o}]=4\lambda B_{o}+2C_{o},  \tag{4.11c}
\end{equation}
\begin{equation}
E_{o}=-4\lambda C_{o},  \tag{4.11d}
\end{equation}
where $\lambda $ is an arbitrary parameter.

We have seen that limiting ourselves to starting from the closed algebra
(4.11a)-(4.11d), namely taking $A=0,B=B_{o},C=C_{o},E=E_{o},$ only trivial
non-local symmetries arise. Thus, as a possible way out we employ a
procedure similar to that applied in Section 3 for the case of Dym equation.
In other words, we search a realization of the prolongation algebra
(4.3a)-(4.3h) such that 
\begin{equation}
A=0,B=B_{o},  \tag{4.12}
\end{equation}

and

\begin{equation}
C=C_{o}+\mu C_{1},  \tag{4.13}
\end{equation}

where $\mu =\mu (y)$ is a function of the pseudopotential to be determined, $%
B_{o},C_{o},D_{o}$ satisfy the relations (4.11a)-(4.11c), and the operator $%
C_{1}\neq 0$ is supposed to obey the commutation rules

\begin{equation}
\lbrack C_{1},B_{o}]=[C_{1},C_{o}]=[C_{1},D_{o}]=0.  \tag{4.14}
\end{equation}

\smallskip For simplicity, we take $E_{1}=C_{1}$ .

Now let us introduce the operator

\begin{equation}
E=E_{o}+\nu E_{1},  \tag{4.15}
\end{equation}

where $E_{o}$ is given by (4.11d) and $\nu $ is a function depending on the
pseudopotential. Let us assume also that the relations

\smallskip

\begin{equation}
\lbrack E_{1},B_{o}]=[E_{1},C_{o}]=[E_{1},D_{o}]=0,  \tag{4.16}
\end{equation}

\begin{equation}
E_{1}\varphi =E_{1}\mu =0,  \tag{4.17}
\end{equation}

are valid.

\smallskip By dealing with the commutation rule (4.9a), it is natural to put

\smallskip

\begin{equation}
\eta =\gamma \mu B_{o}+\eta _{1},  \tag{4.18}
\end{equation}

\smallskip

where $\gamma $ is a constant, and the operator $\eta _{1}$ is chosen in
such a way that

\begin{equation}
\lbrack \eta _{1},B_{o}]=[\eta _{1},C_{o}]=[\eta _{1},D_{o}]=0.  \tag{4.19}
\end{equation}

In order to obtain the infinitesimal generator of the non-local symmetries,
we need to know the field $\varphi $ and the operator $\eta \;$(see (4.10)).
To this aim, first we shall exploit a specific realization of the algebra
(4.11a)-(4.11c). Precisely, let us consider

\smallskip

\begin{equation}
B_{o}=-\partial _{y_{1}},  \tag{4.20a}
\end{equation}

\begin{equation}
C_{o}=(\lambda -y_{1}^{2})\partial _{y_{1}}+y_{1}\partial _{y_{2}}, 
\tag{4.20b}
\end{equation}

\begin{equation}
D_{o}=-2y_{1}\partial _{y_{1}}+\partial _{y_{2}}.  \tag{4.20c}
\end{equation}

Second, we make the hypothesis that the field $\varphi (y)$ and the
functions $\mu (y)$ and $\nu (y)$ depend on a pseudopotential vector having
at least three components, say $y\equiv (y_{1},y_{2},y_{3}).$ Consequently,
in this context it is reliable to suppose that the operators $C_{1}$ and $%
\eta _{1},$ present in (4.13) and (4.18), respectively, take the form

\smallskip 
\begin{equation}
C_{1}=\varphi _{1}(y)\partial _{y_{1}}+\varphi _{2}(y)\partial
_{y_{2}}+\varphi _{3}(y)\partial _{y_{3}},  \tag{4.21}
\end{equation}

and

\begin{equation}
\eta _{1}=X_{1}(y)\partial _{y_{1}}+X_{2}(y)\partial
_{y_{2}}+X_{3}(y)\partial _{y_{3}},  \tag{4.22}
\end{equation}
where $\varphi _{j}(y)$ and $X(y)$ are functions to be determined.

Starting from the prolongation algebra (4.3a)-(4.3h), and keeping in mind
the previous assumptions and the algebraic constraints (4.6a)-(4.6f) and
(4.9a)-(4.9f), in analogy with the case of the Dym equation, we can deduce
another set of constraints which we omit for brevity.

By using all these relations, after some manipulations we have obtained the
following results :

\begin{equation}
\mu =\mu _{1}(y_{3})e^{2y_{2}},  \tag{4.23}
\end{equation}

\begin{equation}
\varphi =-\alpha \mu _{1}(y_{3})y_{1}e^{2y_{2}},  \tag{4.24}
\end{equation}

\begin{equation}
\nu =4\mu _{1}(y_{3})y_{1}^{2}e^{2y_{2}}+\nu _{1}(y_{2},y_{3}),  \tag{4.25}
\end{equation}

\begin{equation}
C_{1}=\varphi _{2}(y_{3})\partial _{y_{2}}+\varphi _{3}(y_{3})\partial
_{y_{3}},  \tag{4.26}
\end{equation}

\begin{equation}
\eta _{1}=X_{2}(y_{3})\partial _{y_{2}}+X_{3}(y_{3})\partial _{y_{3}}, 
\tag{4.27}
\end{equation}

\smallskip

where the functions $\mu _{1}=\mu _{1}(y_{3}),\nu _{1}=\nu
_{1}(y_{2},y_{3}),\varphi _{2}=\varphi _{2}(y_{3}),\varphi _{3}=\varphi
_{3}(y_{3}),$

$X_{2}=X_{2}(y_{3}),X_{3}=X_{3}(y_{3})$ \ fulfil the system of linear
differential equations

\smallskip

\begin{equation}
2\varphi _{2}\mu _{1}+\varphi _{3}\mu _{1y_{3}}=0,  \tag{4.28a}
\end{equation}

\begin{equation}
-\varphi _{2y_{3}}X_{3}+\varphi _{3}X_{2y_{3}}=\frac{\alpha }{4}+2\varphi
_{2}X_{2}+\varphi _{2}X_{3}\frac{1}{\mu _{1}}\mu _{1y_{3}},  \tag{4.28b}
\end{equation}

\begin{equation}
\varphi _{3}X_{3y_{3}}-\varphi _{3y_{3}}X_{3}=\varphi _{3}X_{3}\frac{1}{\mu
_{1}}\mu _{1y_{3}}+2\varphi _{3}X_{2},  \tag{4.28c}
\end{equation}

\begin{equation}
X_{2}\nu _{1y_{2}}+X_{3}\nu _{1y_{3}}=2X_{2}\nu _{1}+X_{3}\nu _{1}\frac{1}{%
\mu _{1}}\mu _{1y_{3}},  \tag{4.28d}
\end{equation}
\begin{equation}
16\lambda \mu _{1}e^{2y_{2}}+\nu _{1y_{2}}=0,  \tag{4.28e}
\end{equation}

\begin{equation}
\phi _{2}\nu _{1y_{2}}+\phi _{3}\nu _{1y_{3}}=0.  \tag{4.28f}
\end{equation}

\smallskip

This system can be handled by means of the procedure used for the Dym
equation.

In doing so, let us explicit the infinitesimal generator (4.10), which reads

\begin{equation}
V_{NL}=-\alpha y_{1}\mu _{1}(y_{3})e^{2y_{2}}\partial _{u}+\frac{\alpha }{4}%
\mu _{1}(y_{3})e^{2y_{2}}\partial _{y_{1}}+X_{2}(y_{3})\partial
_{y_{2}}+X_{3}(y_{3})\partial _{y_{3}}  \tag{4.29}
\end{equation}

Then, from the the group transformations corresponding to the operator
(4.29) we easily find

\begin{equation}
\stackrel{\sim }{u}=-2(\stackrel{\sim }{y}_{1}^{2}-y_{1}^{2})+u.  \tag{4.30}
\end{equation}

\smallskip

In our scheme, if we choose, for instance,$\;u=0,\;$the prolongation
equations (4.2a) and (4.2b) become

\begin{equation}
y_{1x}=\lambda -y_{1}^{2},  \tag{4.31a}
\end{equation}

\begin{equation}
y_{1t}=-4\lambda ^{2}+4\lambda y_{1}^{2},  \tag{4.31b}
\end{equation}

\begin{equation}
y_{2x}=y_{1}+\varphi _{2}(y_{3})\mu _{1}(\stackrel{}{y_{3}})e^{2y_{2}}, 
\tag{4.31c}
\end{equation}

\begin{equation}
y_{2t}=-4\lambda y_{1}+(4y_{1}^{2}-8\lambda )\varphi _{2}(y_{3})\mu _{1}(%
\stackrel{}{y_{3}})e^{2y_{2}},  \tag{4.31d}
\end{equation}

\begin{equation}
y_{3x}=\varphi _{3}(y_{3})\mu _{1}(\stackrel{}{y_{3}})e^{2y_{2}}, 
\tag{4.31e}
\end{equation}

\begin{equation}
y_{3t}=(4y_{1}^{2}-8\lambda )\varphi _{3}(y_{3})\mu _{1}(\stackrel{}{y_{3}}%
)e^{2y_{2}}.  \tag{4.31f}
\end{equation}

Then, for $\lambda >0$ , from Eqs. (4.31a) and (4.31b) we obtain

\begin{equation}
y_{1}=\sqrt{\lambda }\frac{e^{\sqrt{\lambda }\xi }-ae^{-\sqrt{\lambda }\xi }%
}{e^{\sqrt{\lambda }\xi }+ae^{-\sqrt{\lambda }\xi }},  \tag{4.32}
\end{equation}

where $\xi =x-4\lambda t$ and $a$ is a constant.

By integrating Eqs. (4.31c)-(4.31f), from Eq.(4.30) we get (for $u=0)$

\begin{eqnarray}
\stackrel{\sim }{u} &=&\frac{2\lambda (e^{\sqrt{\lambda }\xi }-ae^{-\sqrt{%
\lambda }\xi })^{2}}{(e^{\sqrt{\lambda }\xi }+ae^{-\sqrt{\lambda }\xi })^{2}}%
-2\{\sqrt{\lambda }\frac{(e^{\sqrt{\lambda }\xi }-ae^{-\sqrt{\lambda }\xi
})^{{}}}{(e^{\sqrt{\lambda }\xi }+ae^{-\sqrt{\lambda }\xi })}+  \tag{4.33} \\
&&\alpha \varepsilon \frac{(e^{\sqrt{\lambda }\xi }+ae^{-\sqrt{\lambda }\xi
})^{2}}{4-\varepsilon \alpha [2a(x-12\lambda t)+\frac{1}{2\sqrt{\lambda }}%
(e^{2\sqrt{\lambda }\xi }-a^{2}e^{-2\sqrt{\lambda }\xi })]}\}^{2}.  \nonumber
\end{eqnarray}
\ 

We notice that since $\stackrel{\sim }{x}=x,\stackrel{\sim }{t}=t,$ this
solution is {\it explicit} and contains the well-known soliton solution [10]

\begin{equation}
\stackrel{\sim }{u}=2\lambda \sec h^{2}(\sqrt{\lambda }\xi +\frac{1}{2}\ln 
\frac{\mid \varepsilon \mid }{8\sqrt{\lambda }}),  \tag{4.34}
\end{equation}

which emerges for $a=0$ and $\alpha =-1.$

Another solution to the KdV equation can be obtained in correspondence of
the choice $\lambda <0.$ This is quoted in Appendix B.

\section{Conclusions}

We have developed a procedure to obtain non-local symmetries of the
pseudopotential type of nonlinear field equations whose prolongation algebra 
$L$ is incomplete. We have considered two case studies: the Dym and the KdV
equation, respectively. For both equations, first we have found a
finite-dimensional subalgebra (quotient algebra) ${\cal L}_{o}$ of the
related (incomplete) prolongation algebra $L$. Then, we have tried to use$%
{\cal \ }{\cal L}_{o}$ to look for non-local symmetries.

Unfortunately, through ${\cal L}_{o}$ only trivial symmetries emerge.
Consequently, we have ''extended'' the subalgebra ${\cal L}_{o}$ by
introducing new operators to be determined by the requirement that the
commutation relations defining $L$ and the constraints involved by the
infinitesimal transformations for the non-local symmetries are satisfied.
The determination of these new operators is crucial, since they appear in
the generator of the non-local symmetries. For the two equations under
investigation this task has been successful. From the generator of the
non-local symmetries, expressed in terms of pseudopotential variables, one
can write the corresponding group transformations which enable us to yield
exact solutions of the Dym and the KdV equations. Some of these solutions
are well-known. Notwithstanding, they serve as paradigms to probe and to
illustrate the potentiality of our algebraic approach.

As we can argue from the results achieved on the above-mentioned
applications, our method could be exploited to treat other nonlinear field
equations admitting nontrivial prolongations. But some aspects of the method
remain to be elucidated, and only the accumulation of cases could indicate
the appropriate way of implementation. To be precise, for instance we remark
that a basic role is played by the realization of the ''extended'' algebra
in terms of vector fields depending on pseudopotential variables. In our
calculations, we have chosen simple but{\it \ }nonlinear{\it \ }realizations
(of the polynomial type). Different algebraic realizations (say, polynomial
realizations in higher dimensions or realizations which are not of the
polynomial type) might produce different non-local symmetries and,
correspondingly, different solutions to the original equations. Anyway, this
question is to be explored. Another interesting attempt which deserves to be
made is the use of an infinite-dimensional realization (of the prolongation
algebra $L$) of the Kac-Moody type. Could this kind of realization have a
significant role in the search of non-local symmetries of the
pseudopotential type? Finally, we point out that an important problem is the
extension of our approach to nonlinear field equations in more than 1+1
dimensions. However, due to the fact that at present only a few applications
in higher dimensions have been carried out within the prolongation scheme
[14], this programme is strictly connected with a possible revival of
interest in the extension of the prolongation studies.

\smallskip

\section{Appendix A}

\smallskip

Here we sketch the calculation to find the solution to the Dym equation
(3.1) corresponding to the choice $\lambda <0$ in the Riccati equations
(3.51a) and (3.51b). These give

\smallskip

\begin{equation}
y_{1}=\sqrt{\mid \lambda \mid }\frac{\cos \sqrt{\mid \lambda \mid }\theta
-b\sin \sqrt{\mid \lambda \mid }\theta }{\sin \sqrt{\mid \lambda \mid }%
\theta +b\cos \sqrt{\mid \lambda \mid }\theta },  \tag{A1}
\end{equation}
where $\theta =x+4\mid \lambda \mid t$ and $b$ is a constant.

This formula is the analogous of (3.53). Carrying out the same type of
calculations leading to (3.56), we obtain the transformation

\smallskip

$\stackrel{\sim }{x}=x-[\varepsilon \chi (\sin \sqrt{\mid \lambda \mid }%
\theta +b\cos \sqrt{\mid \lambda \mid }\theta )^{2}]\times $

\smallskip $\{1+\frac{\varepsilon \chi \mid \lambda \mid }{4}[%
2(b^{2}+1)(x+12\mid \lambda \mid t)+\frac{(b^{2}-1)\sin 2\sqrt{\mid \lambda
\mid }\theta -2b\cos 2\sqrt{\mid \lambda \mid }\theta }{\sqrt{\mid \lambda
\mid }}]$

$+\varepsilon \chi \mid \lambda \mid (\cos \sqrt{\mid \lambda \mid }\theta
-b\sin \sqrt{\mid \lambda \mid }\theta )(\sin \sqrt{\mid \lambda \mid }%
\theta +b\cos \sqrt{\mid \lambda \mid }\theta )\}^{-1},$

\begin{equation}
\tag{A2}
\end{equation}

where $\chi =\pm 1.$

\smallskip

If $x=\gamma (\stackrel{\sim }{x},\stackrel{\sim }{t})$ indicates formally
the inverse of the expression (A2), keeping in mind (A1) from (3.52) we have

\smallskip

$\stackrel{\sim }{u}=-\{1+\frac{\varepsilon \chi \mid \lambda \mid }{4}[%
2(b^{2}+1)(\gamma (\stackrel{\sim }{x},\stackrel{\sim }{t})+12\mid \lambda
\mid \stackrel{\sim }{t})+\frac{(b^{2}-1)\sin 2\sqrt{\mid \lambda \mid }%
\stackrel{\sim }{\theta }-2b\cos 2\sqrt{\mid \lambda \mid }\stackrel{\sim }{%
\theta }}{\sqrt{\mid \lambda \mid }}]\}\times $

$\{1+\frac{\varepsilon \chi \mid \lambda \mid }{4}[2(b^{2}+1)(\gamma (%
\stackrel{\sim }{x},\stackrel{\sim }{t})+12\mid \lambda \mid \stackrel{\sim 
}{t})+\frac{(b^{2}-1)\sin 2\sqrt{\mid \lambda \mid }\stackrel{\sim }{\theta }%
-2b\cos 2\sqrt{\mid \lambda \mid }\stackrel{\sim }{\theta }}{\sqrt{\mid
\lambda \mid }}]$

$+\varepsilon \chi \mid \lambda \mid (\cos \sqrt{\mid \lambda \mid }%
\stackrel{\sim }{\theta }-b\sin \sqrt{\mid \lambda \mid }\stackrel{\sim }{%
\theta })(\sin \sqrt{\mid \lambda \mid }\stackrel{\sim }{\theta }+b\cos 
\sqrt{\mid \lambda \mid }\stackrel{\sim }{\theta })\}^{-1},$

\begin{equation}
\tag{A3}
\end{equation}

where $\stackrel{\sim }{\theta }=$ $\gamma (\stackrel{\sim }{x},\stackrel{%
\sim }{t})+4\mid \lambda \mid \stackrel{\sim }{t}.$

\smallskip

\section{Appendix B}

\smallskip By choosing $\lambda <0$, Eqs. (4.31a) and (4.31b) give rise to
the solution expressed by (A1). Following the same procedure adopted to find
the solution (4.33), we obtain

\smallskip

$\stackrel{\sim }{u}=[-4\varepsilon \alpha \sqrt{\mid \lambda \mid }(\cos 
\sqrt{\mid \lambda \mid }\theta -b\sin \sqrt{\mid \lambda \mid }\theta
)(\sin \sqrt{\mid \lambda \mid }\theta +b\cos \sqrt{\mid \lambda \mid }%
\theta )]\times \{4-\frac{\varepsilon \alpha }{4}[2(1+b^{2})\theta $

$+\frac{(b^{2}-1)\sin 2\sqrt{\mid \lambda \mid }\theta -2b\cos 2\sqrt{\mid
\lambda \mid }\theta }{\sqrt{\mid \lambda \mid }}+16\sqrt{\mid \lambda \mid }%
(1+b^{2})t]\}^{-1}-2\varepsilon ^{2}\alpha ^{2}(\sin \sqrt{\mid \lambda \mid 
}\theta +b\cos \sqrt{\mid \lambda \mid }\theta )^{4}\times $

$\{4-\frac{\varepsilon \alpha }{4}[2(1+b^{2})(x+4\mid \lambda \mid t)+\frac{%
(b^{2}-1)\sin 2\sqrt{\mid \lambda \mid }\theta -2b\cos 2\sqrt{\mid \lambda
\mid }\theta }{\sqrt{\mid \lambda \mid }}+16\sqrt{\mid \lambda \mid }%
(1+b^{2})t]\}^{-2},$

\begin{equation}
\tag{B1}
\end{equation}

with $\theta =x+4\mid \lambda \mid t,$ $\alpha \in R.$

\vspace{1.2cm} \noindent {\Large {\bf {Acknowledgements}}} \vspace{0.7cm}

\noindent Support from MURST of Italy and INFN-Sezione di Lecce is
gratefully acknowledged.

\end{document}